\documentclass[aps,prl,twocolumn,showkeys,superscriptaddress]{revtex4-2}

\usepackage{placeins}                   
\usepackage[artemisia]{textgreek}       
\usepackage{amssymb}                    
\usepackage{amsmath} 				    
\usepackage{latexsym}                   
\usepackage{nicefrac}                   

\usepackage{graphicx}                   
\usepackage{epsfig}                     
\usepackage{epstopdf}                   
\usepackage{dcolumn}                    

\usepackage{lipsum}                     
\usepackage{color}                      

\usepackage{subfigure}                  
\usepackage{blindtext}                  
\usepackage[ulem=normalem]{changes}     
\usepackage{float}                      
\usepackage{array}                      
\usepackage{soul}                       
\usepackage[T1]{fontenc}                
\usepackage[hidelinks]{hyperref}        

\usepackage[separate-uncertainty=true,multi-part-units=single]{siunitx}
\DeclareSIUnit\sq{\ensuremath{\Box}}
\DeclareSIUnit\bar{bar}
\DeclareSIUnit\angstrom{\text{Å}}
\DeclareSIUnit{\atpercent}{at\%}

\newcolumntype{P}[1]{>{\centering\arraybackslash}p{#1}}
\hypersetup{
    colorlinks,
    linkcolor={red!50!black},
    citecolor={blue!50!black},
    urlcolor={blue!80!black}
}

\newcolumntype{L}[1]{>{\raggedright\arraybackslash}p{#1}}
\newcolumntype{C}[1]{>{\centering\arraybackslash}p{#1}}
\newcolumntype{R}[1]{>{\raggedleft\arraybackslash}p{#1}}

\begin{document}

\title{Development of Nb-GaAs based superconductor semiconductor hybrid platform by combining in-situ dc magnetron sputtering and molecular beam epitaxy}

\author{Clemens Todt}
\affiliation{Solid State Physics Laboratory, ETH Z\"urich, CH-8093 Z\"urich, Switzerland}
\author{Sjoerd Telkamp}
\affiliation{Solid State Physics Laboratory, ETH Z\"urich, CH-8093 Z\"urich, Switzerland}
\author{Filip Krizek}
\affiliation{Solid State Physics Laboratory, ETH Z\"urich, CH-8093 Z\"urich, Switzerland}
\affiliation{IBM Research Europe - Zurich, 8803 Rüschlikon, Switzerland}
\affiliation{Institute of Physics, Czech Academy of Sciences, 162 00 Prague, Czech Republic}
\author{Christian Reichl}
\affiliation{Solid State Physics Laboratory, ETH Z\"urich, CH-8093 Z\"urich, Switzerland}
\author{Mihai Gabureac}
\affiliation{Solid State Physics Laboratory, ETH Z\"urich, CH-8093 Z\"urich, Switzerland}
\author{R\"udiger Schott}
\affiliation{Solid State Physics Laboratory, ETH Z\"urich, CH-8093 Z\"urich, Switzerland}
\author{Erik Cheah}
\affiliation{Solid State Physics Laboratory, ETH Z\"urich, CH-8093 Z\"urich, Switzerland}
\author{Peng Zeng}
\affiliation{ETH Zürich, The Scientific Center for Optical and Electron Microscopy (ScopeM), CH 8093 Zürich, Switzerland.}
\author{Thomas Weber}
\affiliation{X-ray Platform, Department of Materials, ETH Zürich, Vladimir-Prelog-Weg 5–10, 8093 Zürich, Switzerland.}
\author{Arnold M\"uller}
\affiliation{Laboratory of Ion Beam Physics, ETH Zurich, Schafmattstrasse 20, CH-8093 Zurich, Switzerland.}
\author{Christof Vockenhuber}
\affiliation{Laboratory of Ion Beam Physics, ETH Zurich, Schafmattstrasse 20, CH-8093 Zurich, Switzerland.}
\author{Mohsen Bahrami Panah}
\affiliation{Solid State Physics Laboratory, ETH Z\"urich, CH-8093 Z\"urich, Switzerland}
\author{Werner Wegscheider}
\affiliation{Solid State Physics Laboratory, ETH Z\"urich, CH-8093 Z\"urich, Switzerland}

\date{\today}

\begin{abstract}
We present Nb thin films deposited in-situ on GaAs by combining molecular beam epitaxy and magnetron sputtering within an ultra-high vacuum cluster. Nb films deposited at varying power, and a reference film from a commercial system, are compared. The results show clear variation between the in-situ and ex-situ deposition which we relate to differences in magnetron sputtering conditions and chamber geometry. The Nb films have critical temperatures of around $\SI{9}{\kelvin}$ and critical perpendicular magnetic fields of up to $B_{c2} = \SI{1.4}{\tesla}$ at $\SI{4.2}{\kelvin}$. From STEM images of the GaAs-Nb interface we find the formation of an amorphous interlayer between the GaAs and the Nb for both the ex-situ and in-situ deposited material. 

\end{abstract}
\pacs{}
\keywords{niobium, dc magnetron sputtering, semiconductor superconductor hybrid materials, in-situ superconductor growth}
\maketitle


\section{Introduction}
Superconductor (SC) semiconductor (SE) hybrid (SSH) devices have re-emerged \cite{1985takayanagi,1990bending,1992nguyen} fueled by the hope of finding anyons in solid state systems and their subsequent application for fault tolerant quantum computing \cite{2003kitaev,2008nayak,2011alicea,2017karzig,2020aguado}. Most notably the search for the Majorana Fermion in solid state systems has attracted attention \cite{2012das,2012mourik}. This promoted numerous experiments in Andreev interaction with Quantum Hall states \cite{2015wan,2017lee,2022gul} and topological superconductivity \cite{2017sato,2019fornieri}.

The achievement of epitaxial growth of thin film Al on III-V SEs \cite{1999Pilkington,2015krogstrup} sparked experiments in SSH devices \cite{2015chang,2016kjaergaard,2016shabani,2017drachmann,2017kjaergaard,2017suominen,2018casparis,2018bottcher,2018Ofarrell,2019lee,2019mayer,2021drachmann,2020nichele,2020whiticar,2022haxell,2022haxella,2022banerjee}. The crucial element of material synthesis is the in-situ deposition, enabling an undisturbed SC-SE combination, crucial for a transparent interface to electron transport \cite{2015krogstrup}. The formation of sub-gap states and a electrostatic barrier, degrading the performance of the hybrid system, is associated to the surface oxide, formed when the SC is deposited ex-situ \cite{2015chang}. 

Therefore, nanowires and two dimensional electron systems (2DES) based on InAs and InSb with an epitaxial Al layer have become the established material platform exhibiting a pronounced proximity effect \cite{2015krogstrup,2015chang,2017gusken}. Furthermore, Al is typically available in molecular beam epitaxy (MBE) systems owing to its use in III-V semiconductor growth. 

The superconducting properties of epitaxial Al films limit the temperature and magnetic field range of Al-based SSH experiments to $T_c$ of around $\SI{1.6}{\kelvin}$ at film thicknesses between $\SI{5}{\nano\m}$ and $\SI{10}{\nano\m}$ \cite{2016kjaergaard,2016shabani,2018bottcher,2019lee,2019mayer}. The reported perpendicular critical fields $B_{c2}$ range from $\SI{30}{\milli\tesla}$ \cite{2017drachmann} up to $\SI{164}{\milli\tesla}$ \cite{2019mayer} at dilution fridge temperatures.

The search for an alternative to Al is the subject of a multitude of recent studies \cite{2017gusken,2020carrad,2021kanne,2021pendharkar,2021dang,2021perla,2022kousar}. A wide range of elemental superconductors has been deposited onto nanowires including Pb \cite{2020khan,2021kanne}, In \cite{2021bjergfelt}, Ta \cite{2020carrad}, V \cite{2019bjergfelt} and Sn \cite{2021pendharkar,2020khan}. Nb is of particular interest \cite{2017gusken,2021perla,2020carrad} as it has the highest bulk critical temperature and magnetic field of all the elemental SCs \cite{2005buzeaa}. Pb appears to be the best alternative so far \cite{2020khan,2021kanne} owing to its favourable lattice match to InAs \cite{2017jelver} and relatively high Tc \cite{2021kanne}. 

Exciting research proposals \cite{2012vanheck,2018obrien,2021fatemi} call for building increasingly complex SSH devices and networks. In this application lithographically patterned 2DES-SC SSH represent a promising approach \cite{2018lutchyn}. The 2DES in InAs \cite{2022haxell} and InSb \cite{2022lei} can be grown to reasonably high mobilities but lack far behind 2DES based on GaAs \cite{2021kulah}. The drawback of GaAs is the \hbox{$\Phi_B > \SI{0.7}{\electronvolt}$} Schottky barrier \cite{1988mclean} which is expected to suppress the proximity effect \cite{1982blonder}. Nonetheless induced superconducting gaps in bulk n-GaAs employing in-situ deposited Al have been measured \cite{1996taboryski,1997kutchinsky}. In this context our recently developed shallow GaAs 2DESs \cite{2021kulah} are posing an interesting unexplored potential for SSHs.    

The interaction between the SE and SC is not limited to the proximity effect. A type-II superconductor can shape the magnetic field in the SE underneath via its vortices \cite{2007weeks} forming the basis of exciting experimental proposals \cite{2009rosenberg,2016zocher,2022okugawa}. Vortex interaction mediated experiments have been previously attempted \cite{1990bending,1991kruithof} most notably by Geim et al. \cite{1992geim,1992geima}. The authors investigated Pb on a GaAs 2DES and concluded that a SC with a small vortex is needed together with a low electron density 2DES as close as possible to the surface \cite{1992geimb}. These requirements are rooted in a geometrical argument, considering the size of the magnetic field variation at the depth of the 2DES versus the size of the quasiparticle in the 2DES as a function of magnetic field. The small vortex size \cite{2004ghenim} can be achieved in Nb and the lowest electron densities have been reached in (Al)GaAs based systems 2DES \cite{1990sajoto}. 

In this work, we present the first results from our chamber for DC magnetron sputtering of SC on our MBE grown III-V SEs without breaking the vacuum. In the initial experiment, we compare Nb deposited in-situ at varying power in the UHV dc magnetron sputtering system and ex-situ deposited in a commercial system (AJA Int.). The samples are of high purity but display significant differences in surface roughness and crystallite orientation which can be related to the growth regime.

We compare the superconducting properties of the Nb film, by investigating the resistive transition as a function of temperature and magnetic field. STEM images of the Nb-GaAs interface reveal an amorphous interlayer at the interface for both the in-situ and ex-situ depositions. 

\section{MBE and Magnetron Sputtering Cluster}
\begin{figure}[h]
    \centering
    \includegraphics[width = \linewidth]{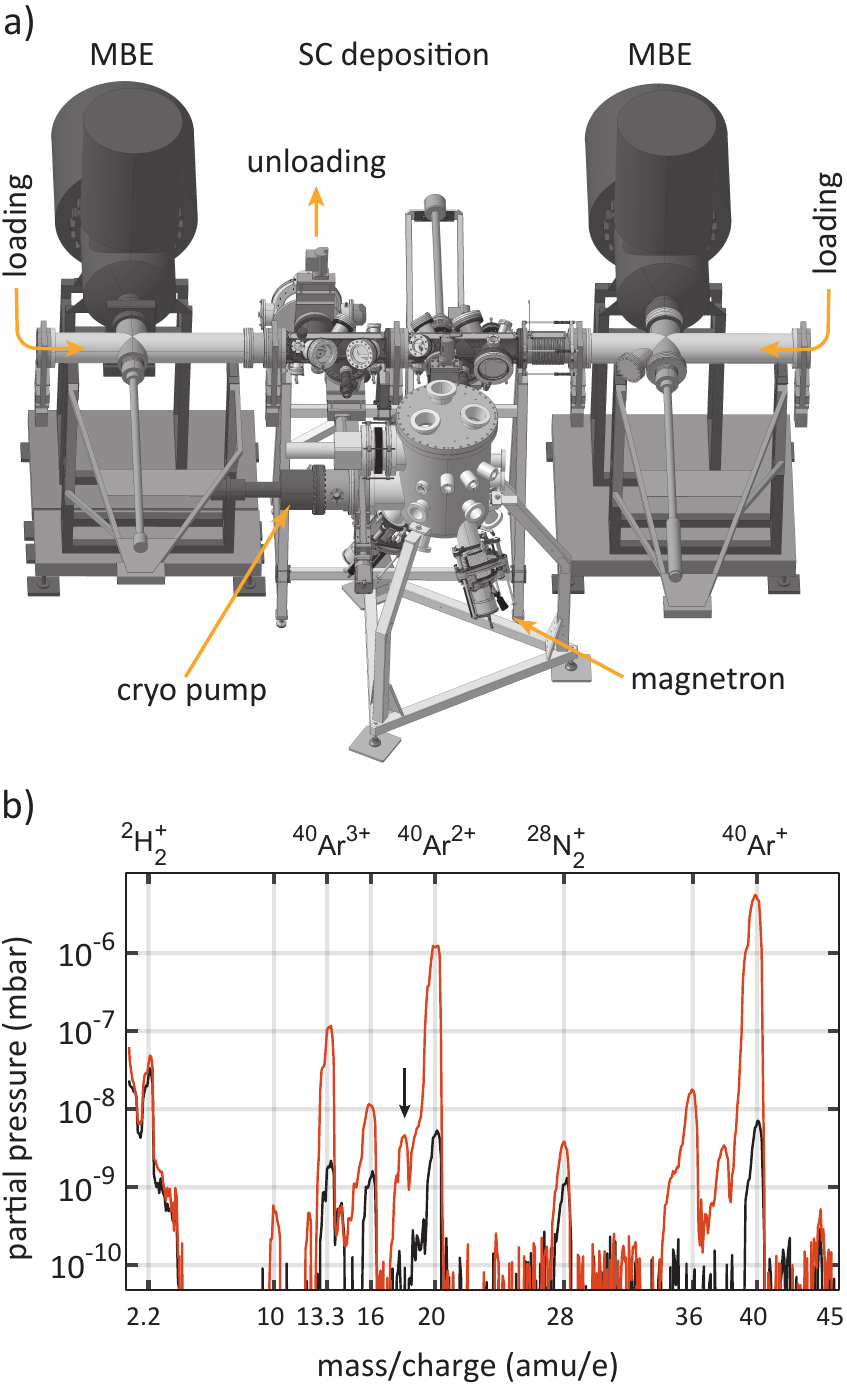}
    \caption{a) Layout of the UHV cluster, consisting of two MBE chambers used for semiconductor growth and the magnetron sputtering chamber for superconductor deposition. b) Mass spectrum of the superconductor deposition chamber. The red line indicates a measurement 18hrs after a Nb deposition while the black line is a measurement taken after pumping the system for a week. The arrow indicates the minute water peak that appears after deposition.}
    \label{fig:chamber}
\end{figure}

The layout of the UHV cluster, consisting of two molecular beam epitaxy (MBE) machines and the SC deposition system, is presented in fig.\ref{fig:chamber} a). The first MBE chamber is optimized for high mobility 2DES in (Al)GaAs \cite{2016berl,2019meyer,2020roosli,2021scharnetzky,2021kulah} while the other covers a wider range of \hbox{III-V} materials based on As and Sb \cite{2019lei,2019karalic,2019shibata,2020shibata,2020lei,2022lei,2022leia}. The UHV magnetron sputtering chamber is connected via a UHV tunnel to enable in-situ deposition of SCs on MBE grown SEs as well as preventing contamination of the MBE systems.

The incorporation of oxygen in superconducting films is generally believed to have a detrimental effect on the superconducting properties such as the critical temperature \cite{1974koch}. Therefore, the system was designed to minimize contaminants, specifically the incorporation of oxygen. The UHV magnetron sputtering chamber is supplied with purified gas and solely pumped by a cryo pump in order to obtain elementally clean films, see supplementary for details \cite{2023todt}. After bakeout the system achieved the mass spectra presented at the bottom fig.\ref{fig:chamber} b). The black line represents the pumped state at $p < \SI{1e-10}{\milli\bar}$ while the red line at $p = \SI{1e-9}{\milli\bar}$ was taken 18hrs after the deposition of Nb. The pressure is dominated by peaks from the different ionization states of Ar and its isotopes (36, 38). A peak associated with water three orders of magnitude smaller than Argon can be identified after deposition indicated by the arrow. Continued use of the chamber reduced the water peak below detection limit and therefore we assume that it originated from residual water in the gas lines. 

The kinetic rather than thermal nature of magnetron sputtering offers an alternative path to the evaporation for SC deposition on SEs. The film growth via evaporation is primarily controlled by substrate temperature and rate \cite{1995palmstrom}. In order to produce connected films of low melting point elements such as In, Pb or even Al on SE surfaces, the substrate has to be typically cooled below room temperature \cite{1990sands,2010brillson,2021kanne,2015krogstrup} adding technical complexity. SC with higher melting points such as Ta and Nb can be grown at higher substrate temperatures \cite{2021perla}. However, to evaporate these low vapor pressure metals \cite{1984alcock} they have to be heated to high temperatures causing the chamber to release contamination from the chamber walls and heat up the substrate surface. Magnetron sputtering, on the other hand, is a comparatively cold deposition method \cite{2010mattox}. 

The method appeals additionally with the possibility to grow nitrides, a simple material exchange, wide variety of compounds from mixed targets and co-sputtering as well as a moderate pressure during deposition which limits outgassing \cite{2010mattox}. This opens up the possibility to deposit a wide range of compound materials such as A15 and B1 phase SCs as well as more exotic variants like  MgB$_2$ \cite{2006micunek}. 

\subsection{Sample preparation}
Both the ex-situ and in-situ Nb films were deposited onto $\SI{720}{\nano\m}$ of MBE grown n$^{++}$ GaAs. The ex-situ wafer was removed from the MBE chamber after growth. Before loading it into the AJA magnetron sputtering system the wafer was etched in a 1:1 solution of HCl ($\SI{32}{\percent}$):H$_2$O at temperature until the surface was hydrophilic to remove the oxide. The wafer was then transferred in air to the load lock within 5 min. The in-situ wafer was moved in our UHV tunnel from the MBE chamber to the UHV magnetron sputtering chamber under a residual pressure of $< \SI{5e-9}{\milli\bar}$.

\subsection{Nb depositions}
In order to investigate the possible operating conditions in our sputtering system, a characterisation of the Nb sputtering rate for various power and pressure combinations was made. This exploration served as a starting point to determine which sputtering conditions could be compared to the commercial AJA system and could yield films with good superconducting parameters.

\begin{figure}[h]
    \centering
    \includegraphics[width=\linewidth]{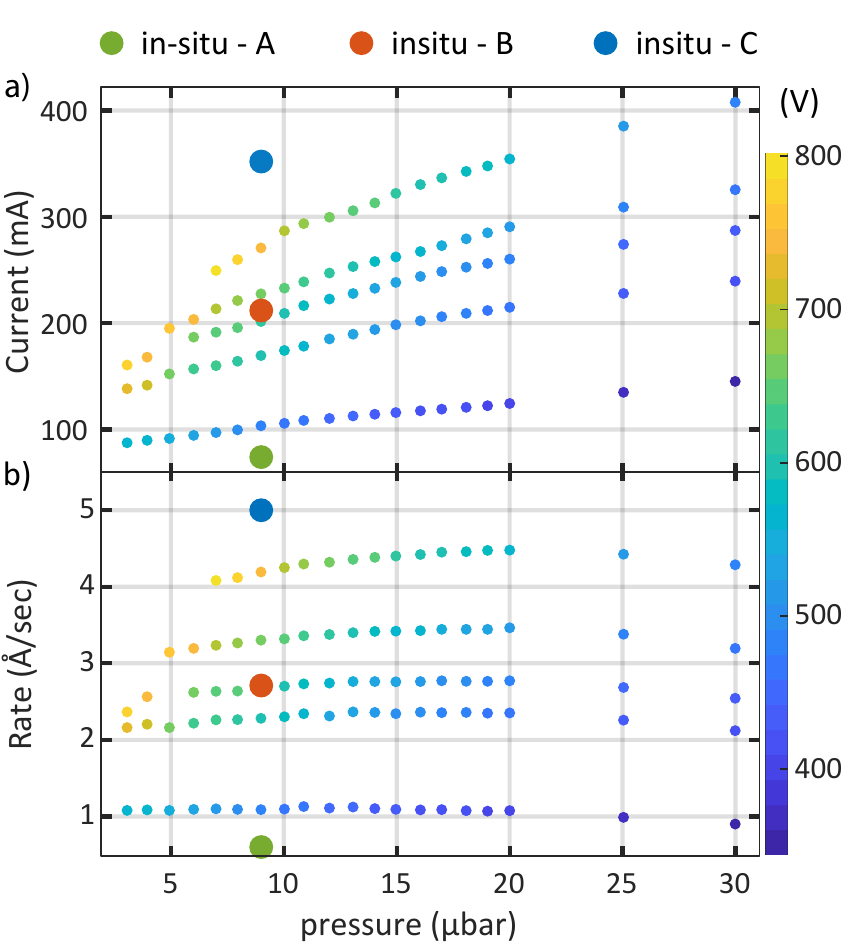}
    \caption{a) Current and b) Nb deposition rate dependence on pressure. The voltage is indicated by the color of the dot and the scale bar. The working conditions for the presented in-situ films are indicated by the larger dots.}
    \label{fig:SputterParameters}
\end{figure}

The dependence of the Nb deposition rate on pressure and set power for our system was investigated using a quartz crystal balance that can be moved into the wafer position and results are presented in fig.\ref{fig:SputterParameters}. The rate increases with pressure up to $\SI{20}{\micro\bar}$ at which point the Nb growth is limited by diffusion from the Ar gas. With increasing pressure the voltage decreases and the current increases as expected from a denser and more conductive plasma. The rate is linearly dependent on the set power at a given pressure.
 
The deposition in our system with $2$ inch UHV magnetrons from Angstrom Sciences is controlled by pressure, power and substrate heating. The parameters used in this study are summarised in table \ref{table:SputterParameters}. The guns are mounted such that we can vary the substrate target distance under a fixed angle of $\SI{32}{\degree}$ and the substrate is not rotated. For this study we chose to fix the distance at the minimum of $\SI{110}{\milli\m}$. The commercial system employs a $4$ inch target $\SI{100}{\milli\m}$ away from the substrate in a planar orientation and a constant substrate rotation. The pressure was chosen such that the pressure-distance product is $\SI{1}{\micro\bar\m}$ for both setups.  

Due to the difference in target size between the systems it is not possible to attain the same rate at the same current and voltage values for both setups. The ex-situ system can attain a low voltage of \SI{214}{\volt} at a high rate while the in-situ machine is limited to $\SI{404}{\volt}$ before the plasma becomes unstable at $\SI{9}{\micro\bar}$. 

\begin{table*}
\centering
\begin{tabular}{L{2cm}                 | C{1.5cm}              |  C{1.5cm}                         |  C{1.5cm}                  | C{1.5cm}          |  C{1.5cm}                               | C{1.5cm}                        ||  C{1.5cm}        } 
                    sample              & substrate             & p                                 & T$_{sub}$                 & voltage           & J                                       & rate                            & R$_{sq}$         \\
                                        &                       & ($\si{\micro \bar}$)              & ($\si{\degreeCelsius}$)   & ($\si{\volt}$)    & ($\si{\milli\ampere\per\cm\squared}$)   & ($\si{\angstrom\per\second}$)   & ($\si{\nano\m}$)   \\
    \hline \hline
                ex-situ                 & GaAs                  & 10                                & RT                        & 214               & 24.2                                    & 5.0                             & 0.95             \\
                in-situ - A             & GaAs                  &  9                                & RT                        & 404               & 3.7                                     & 0.6                             & 1.84             \\
                in-situ - B             & GaAs                  &  9                                & RT                        & 582               & 10.6                                    & 2.7                             & 1.75             \\
                in-situ - C             & GaAs                  &  9                                & RT                        & 708               & 17.4                                    & 5.0                             & 2.03             \\
    \hline
    \cite{2012dobrovolskiy} B2          & Al$_2$O$_3$           & 4                                 & 850                       & 312               & 2.5                                     & 5.0                             & 0.7              \\
    \cite{1992imamura}                  & Si                    & 23                                & RT                        & 270               & 74                                      & 27.3                            &                  \\
\end{tabular}
\caption{comparison of dc sputtering parameters p - pressure, T$_{sub}$ - substrate temperature, voltage, J - current density and rate with the resulting root mean square roughness R$_{sq}$ obtained from AFM measurements in fig.\ref{fig:AFM}. The bottom three lines are comparable films from literature, see text.}
\label{table:SputterParameters}
\end{table*}

\section{Structural and Elemental Analysis}
\subsection{AFM}
\begin{figure}[!h]
    \centering
    \includegraphics[width=\linewidth]{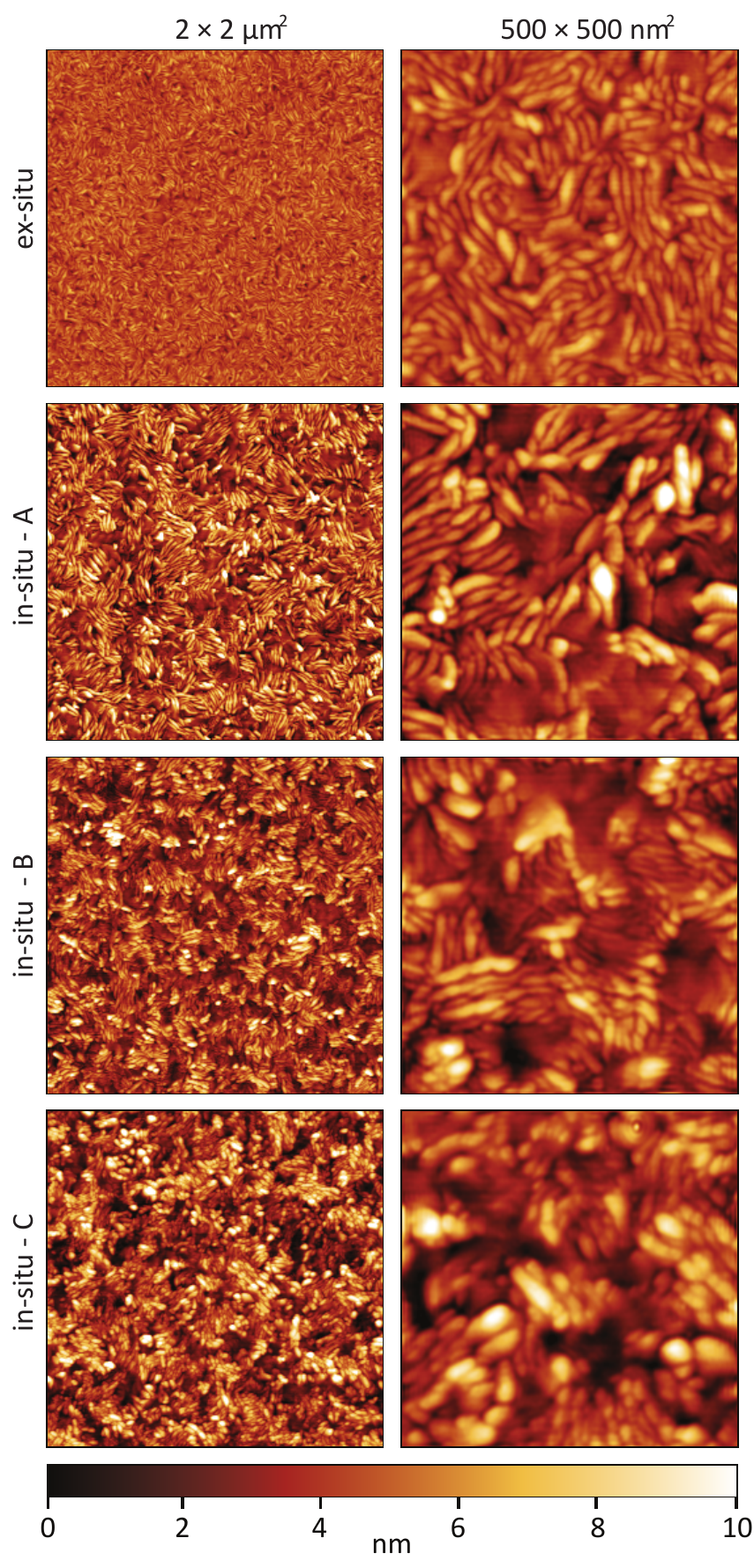}
    \caption{Topography maps of the surface of the in-situ deposited Nb films A,B and C and the ex-situ deposited Nb films. Maps acquired 2x2$\si{\micro\m\squared}$ areas are shown in the left panel and from 500x500$\si{\nano\m\squared}$ ares in the right panels.}
    \label{fig:AFM}
\end{figure}

Fig.\ref{fig:AFM} shows the surface morphologies of the Nb films measured by AFM. All samples have randomly distributed elongated grains, roughly $\SI{100}{\nano\m}$ long and $\SI{20}{\nano\m}$ wide. The elongated grains are not oriented with respect to the substrate or the source. Since randomly oriented elongated Niobium grains are also observed on a silicon substrate by Imamura et. al. \cite{1992imamura}, a direct relationship between this effect and the GaAs substrate seems unlikely.

The root mean square roughness values obtained from the AFM data using Gwyddion \cite{2012necas} are listed in the last column of table \ref{table:SputterParameters}. The in-situ films are distinctively rougher than the ex-situ film with little difference between the in-situ films. 

\subsection{XRD}
\begin{figure}[!h]
    \centering
    \includegraphics[width=\linewidth]{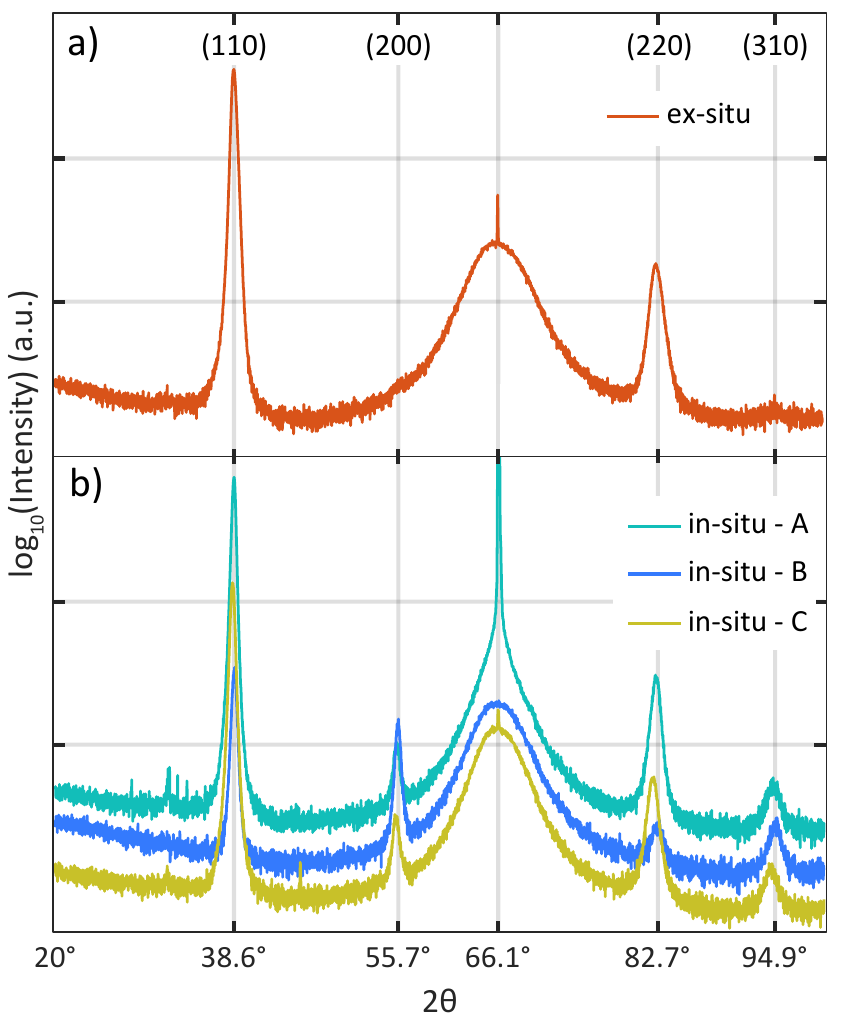}
    \caption{XRD measurements using a PANalytical X’PERT PRO MPD diffractometer in Bragg-Brentano reflection geometry and Cu K$\alpha_1$ radiation. Tails of the GaAs 001 reflection appear as broad background at $\SI{66.1}{\degree}$m while the sharp contributions at this position are from small slightly misoriented substrate crystallites, likely from the cut edge. a) is the ex-situ sample while b) shows traces from the in-situ samples vertically offset for clarity.}
    \label{fig:XRD}
\end{figure}

XRD measurements were performed with a PANalytical X’PERT PRO MPD diffractometer in Bragg-Brentano reflection geometry and Cu K$\alpha_1$ radiation. The measurements of the Nb films are plotted in fig.\ref{fig:XRD}. The data was acquired under a  $\SI{2}{\degree}$ offset of the sample tilt relative to the symmetric geometry to minimize the signal from the GaAs substrate. The signal from the (001) oriented GaAs substrate still appears as broad background centered at $\SI{66.1}{\degree}$, which is coming from thermal diffuse scattering. Apart from the substrate signals, no significant differences are observed between the measurements with and without offset.

Comparison of our XRD data with literature \cite{2016defreitas,2018wilde} shows that our in-situ Nb films are missing the reflections associated with the (211) orientation parallel to the substrate surface \cite{2016defreitas,2018wilde}. The ex-situ film, on the other hand, only shows the 110 and 220 reflections indicating that the crystallites, making up the uniform film, have a preferential orientation of the (100) planes with respect to the substrate. 

The relative peak heights of the 110 family of reflections and the 200 reflection vary between the in-situ samples while the 310 signal appears unchanged. Depending on deposition voltage the orientation distribution either the (110) or (200) oriented crystallites appears to vary. However, it is not a direct trend as the largest variation is observed for the \hbox{in-situ - B sample}.

\subsection{Elemental analysis}
Rutherford Back Scattering (RBS) measurement of the Nb films were undertaken. The RBS was performed using $\SI{2}{\mega\electronvolt}$ Helium ions under a back-scattering angle of $\SI{167.5}{\degree}$. The particle induced X-ray emission (PIXE) from the sample was measured in parallel. No significant contamination of the Nb film could be conclusively detected within the capabilities of RBS and PIXE, see supplementary for details \cite{2023todt}. 

\subsection{Discussion}
The possible origin of the observed structural variations between the in-situ and ex-situ films could be related to different deposition parameters, namely voltage, substrate surface and or geometric differences between the two sputtering systems.

It appears, that the large kinetic energy difference related to different powers used for the in-situ samples do not make a significant difference in their structure that could be identified with the implemented characterization methods. A step up in deposition voltage between the in-situ and ex-situ samples does exist. However, we observed very little structural change when changing the voltage from 404V to 708V for samples in-situ - A,B and C. Given that both systems work at the same pressure distance product, this does not point to the deposition voltage as the root cause for our observed structural differences.

To aid in understanding our findings and bring our deposition conditions into context with those from published literature which have been appended to table \ref{table:SputterParameters}.

Dobrovolskiy et al. \cite{2012dobrovolskiy} have reached the Stranski-Krastanov regime growth regime \cite{2010mattox} and produced epitaxial Nb films at $\SI{850}{\degreeCelsius}$ on Al$_2$O$_3$. The key differences between the reference film by Doborvolskiy et al. and our material is the Al$_2$O$_3$ substrate which not only has a favourable lattice match but also allows for the required high substrate temperature. The high quality clean limit film listed in table \ref{table:SputterParameters} has been grown at the same rate as our ex-situ and in-situ - C samples. In terms of \hbox{voltage} \hbox{current} conditions the reference film is comparable to our in-situ - A sample which suggests that the authors had a smaller substrate target distance to achieve a ten fold higher rate at roughly half the pressure. The roughness increased with elevated growth rate corresponding to a larger voltage which could either be due to the kinetic energy of arriving species or adatoms that didn't have enough time to reach a kink site. 

Imamura et al.\cite{1992imamura} report similarly elongated grains on their Nb films deposited on Si at room temperature. The distance pressure product employed by Imamura et al. is close to ours at $\SI{1.3}{\micro\bar\m}$ which could explain the close resemblance. The authors find that with reducing pressure or equivalently increasing voltage the width of the XRD peaks reduces (therefore crystallite size increases) and the surface becomes smoother. Specifically, the strain went from tensile to compressive at $\SI{270}{\volt}$ which corresponds to the point at which crystallite size and roughness did not change anymore. Our in-situ samples deposited at voltages $>\SI{404}{\volt}$ appeared to follow this in that roughness and crystallite size did not change. 

The random crystallite orientation seen in the XRD in fig.\ref{fig:XRD} on the in-situ material suggests that the Nb does not find a preferential orientation with the substrate. The in-situ material presents the clean GaAs surface reconstruction while the ex-situ material was exposed to air. It is therefore less likely that the XRD findings are not originating from the Nb-GaAs interface. 

The film growth as we understand it has been discussed in detail by Monti et al. \cite{2023monti}. Randomly nucleated crystallites grow in the direction of the facet that incorporates new material at the highest rate. Which facets grow is determined by the surface energy of the specific facet, the adatom mobility on the surface, the direction and rate of the arriving Nb. This will result in the crystallites with favourable orientation to outgrow and terminate neighbouring crystallites until only dendrites of one orientation remain. 

The fact that the ex-situ sample only shows the (110) family of crystal orientations parallel to the substrate surface, could therefore originate from a fast saturation of the termination process. The (110) facets thus grow the fastest and terminate their neighbours faster in the ex-situ system than in the in-situ system. The in-situ samples on other hand have not reached the point at which slower growing crystallites are buried resulting in a rougher surface and more crystallite orientations appearing in the XRD. The (211) oriented crystallites have been buried early on in both systems and do not appear at all in the XRD data.

\section{Nb superconducting properties}
The critical temperature $T_c$ and critical magnetic field $B_{c2}$ of the Nb films limit the measurement range of SSH devices. $B_{c2}$ is the field applied perpendicular to the film at which the resistive transition occurs, termed the upper critical field associated with type-II superconductors \cite{1981kerchner}. Knowledge of the resistive superconducting transition can additionally be used to estimate the coherence length of the Cooper pairs. When compared to the mean free path of the electrons $\ell$ the coherence length determines whether the superconducting film is in the clean or dirty limit \cite{1996tinkham}. 

In the context of proximity induced superconductivity the coherence length appears in the pair breaking parameter indicating that a larger coherence length enhances the pair breaking within the superconductor towards the interface \cite{2003schapers}. 

\begin{figure}[b]
    \centering
    \includegraphics[width = \linewidth]{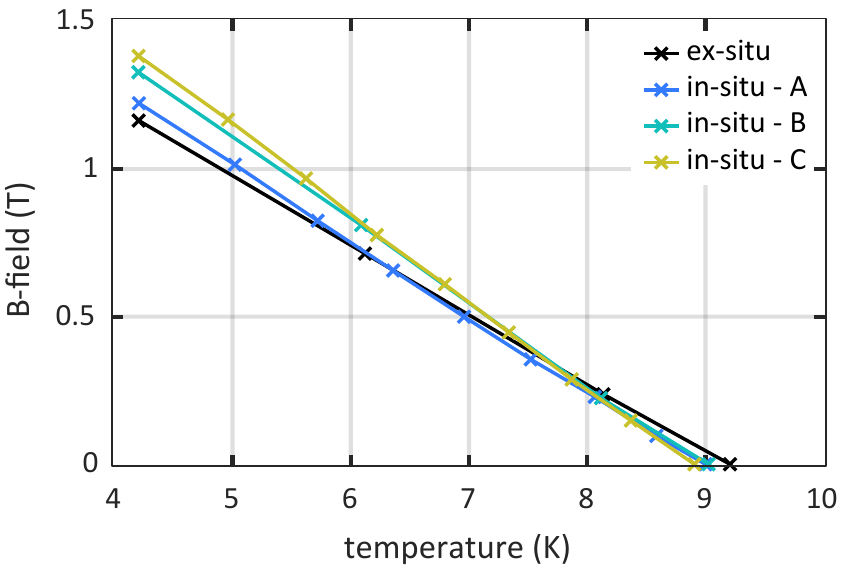}
    \caption{the resistive superconducting transition as a function of temperature and perpendicular magnetic field for the material deposited ex-situ and the films A, B and C deposited In-situ at a rate of $\num{0.6}$, $\num{2.7}$ and $\SI{5}{\angstrom\per\second}$, respectively.}
    \label{fig:TcBc}
\end{figure}

Van der Pauw structures were made and measured using standard lock-in techniques to determine the sheet resistance as a function of temperature and magnetic field $R(T,B)$, see supplementary for details \cite{2023todt}. The result of measuring the resistive transitions at a constant temperature, while sweeping the magnetic field, is given in fig \ref{fig:TcBc}. 

The extracted $B_{c2}(T)$ values are significantly higher than expected for clean limit films which have $B_{c2}(0)$ values of $\SI{1}{\tesla}$ or less \cite{2005peroz,2012dobrovolskiy}. It has been reported that dirty limit films have increased critical fields \cite{2012dobrovolskiy, 2005peroz} while the critical temperature is lower. The large critical field of our films indicates that these are in the dirty limit with the ex-situ sample being closest to the clean limit. The results shown in fig.\ref{fig:TcBc} indicate that, even for room temperature Nb depositions, it is possible to obtain films with less structural defects by changing the sputtering parameters such as power and voltage.

\begin{table}[htb]
\centering
\begin{tabular}{ L{1.5cm}                       | C{0.8cm}       | C{1.2cm}              | C{1cm}                           | C{1.0cm}         | C{1cm}           | C{1cm} } 
  sample                                        & $T_c$          & $\rho$                 & $\ell$ eq.\ref{eq:mfp-Mayadas}  & $B_{c2}(0)$       & $\xi_{GL}(0)$    & $\ell$ eq.\ref{eq:TcBc-dirty} \\
                                                & ($\si{\kelvin}$)& ($\si{\micro\ohm\cm}$)& ($\si{\nano \m}$)               & ($\si{\tesla}$)   & ($\si{\nano\m}$) & ($\si{\nano\m}$)  \\  
\hline \hline
ex-situ                                        & 9.2            & 47.8                   & 0.8                              & 1.48              & 15               & 2.3 \\ 
in-situ - A                                    & 9.0            & 32.7                   & 1.1                              & 1.59              & 14               & 2.1 \\ 
in-situ - B                                    & 9.0            & 43.8                   & 0.8                              & 1.73              & 14               & 1.9 \\ 
in-situ - C                                    & 8.9            & 71.7                   & 0.5                              & 1.83              & 13               & 1.8 \\ 
\hline
\cite{2012dobrovolskiy} B2                     & 9.1            & 0.45                   & 83                               & 0.7               & 22               &     \\
\end{tabular}
\caption{superconducting properties of the Nb films. The film thicknesses are \SI{100}{\nano\m} from QCM rates with the exception of in-situ - B which is $\SI{90}{\nano\m}$. The reference film by Dobrovolskiy et al. \cite{2012dobrovolskiy} is \SI{52}{\nano\m} thick. The normal state resistivity $\rho$ was measured at $\SI{10}{\kelvin}$ just above the critical temperature $T_c$. $\ell$ is the mean free path determined using $\rho$ in eq.\ref{eq:mfp-Mayadas}. $\xi_{GL}(0)$ is obtained when using the GL eq.\ref{eq:Bc2}. The critical perpendicular magnetic field at zero temperature $B_{c2}(0)$ is estimated from the WHH expression in eq.\ref{eq:WHH}. The mean free path $\ell$ the last column arrived at from the GL expression in eq. \ref{eq:TcBc-dirty}. }
\label{table:comparison}
\end{table}

The target film thickness was chosen to be $\SI{100}{\nano\m}$ such that the superconducting critical temperature \cite{2014zaytseva} and critical field \cite{1986quateman} of Nb are not expected to significantly change with variations in thickness \cite{1994minhaj,2020zaytseva}. The thicknesses are determined from QCM deposition rates and time with the in-situ - B sample being slightly thinner at $\SI{90}{\nano\m}$. 

\subsection{Mean free path}
Estimating $\ell$ from the Drude formula \cite{1976ashcroft}
\begin{align}\label{eq:Drude}
    \rho \ell&= \frac{m^*}{e^2}\frac{v_F}{n}
\end{align}
requires knowledge of the normal state resistivity $\rho$, the carrier density $n$ and the effective mass $m^*$ at a temperature just above the superconducting transition. Ideally, each of these parameters is determined from a separate measurement. However, it is common \cite{2005gubin,2009hazra,2012dobrovolskiy} for Nb films to estimate $\ell$ just from $\rho$ using the expression by Mayadas et al. \cite{1972mayadas}, given as
\begin{align}\label{eq:mfp-Mayadas}
    \rho \ell &= \SI{3.72e-6}{\micro \ohm \cm \squared}.
\end{align}

The normal state resistivity of the sample is calculated from the measured resistance at $\SI{10}{\kelvin}$ and zero applied magnetic field using \cite{2006schroder}
\begin{align}\label{eq:rho-simple}
    \rho &= \frac{\pi t}{\ln\left( 2 \right)}R(\SI{10}{\kelvin},\SI{0}{\tesla})
\end{align}
where $t$ is the Nb film thickness determined from the QCM rate and deposition time. The extracted $\ell$ values do compare well with published values \cite{2005gubin,2011leo,2012dobrovolskiy,2018pinto,2020zaytseva} and are listed in fourth column of table \ref{table:comparison}.

\subsection{Ginzburg Landau coherence length}
The Ginzburg- Landau (GL) coherence length $\xi_{GL}(T)$, which denotes the characteristic length scale over which the order parameter varies, is arrived at via the upper critical field
\begin{align}\label{eq:Bc2}
    B_{c2}(T) &= \frac{\phi_o}{2\pi\xi_{GL}^2(T)}.
\end{align}
where $\phi_o = \frac{h}{2e}$ is the magnetic flux quantum in type-II superconductors. 

To obtain a value for $\xi_{GL}(0)$ the zero temperature critical field has to be estimated. The resistive transition $R(T,B)$ from fig.\ref{fig:TcBc} is not linear down to zero temperature and cannot be simply extrapolated. Werthamer, Helfand and Hohenberg (WHH) \cite{1966werthamer} arrived at a relevant theory taking into account non-magnetic impurities, spin paramagnetism and spin-orbit scattering at high fields based on initial results by Maki \cite{1966maki}. The relevant result from WHH has been presented by Gurevich et al. \cite{2003gurevich} as
\begin{align}\label{eq:WHH}
    B_{c2}(0) &= 0.69T_c\left.\frac{dB_{c2}}{dT}\right|_{T = T_c}.
\end{align}
Applying this theory produces the values for $B_{c2}(0)$ and $\xi_{GL}(0)$ presented in table \ref{table:comparison}. The GL coherence length compares well with previously reported values \cite{2012dobrovolskiy,2014ilin,1999wang} and the expected $B_{c2}(0)$ are one order of magnitude larger than what has been achieved with thin Al films \cite{2019mayer}.

\subsection{Bardeen Cooper Schrieffer coherence length}
Although eq.\ref{eq:mfp-Mayadas} is an established method, there are critiques relevant in the context of thin polycrystalline films \cite{1938fuchs,1970mayadas,1987Vancea}. It therefore is warranted to sanity check the consistency of eq.\ref{eq:mfp-Mayadas} with the Ginzburg-Landau-Abrikosov-Gor'kov (GLAG) theory \cite{2016mangin}. It connects the findings of GL and Bardeen-Copper-Schrieffer(BCS) for dirty limit films giving a relation between $\xi_{GL}(T)$ and $\xi_o$ \cite{1996tinkham} near $T_c$ as
\begin{align}\label{eq:xi_GL_BCS}
   \xi_{GL}(T) &= 0.85\sqrt{\ell \xi_o}\left(1 - \frac{T}{T_c} \right)^{-\frac{1}{2}}
\end{align}
where $\xi_o$ is the BCS coherence length at zero temperature. 

The BCS theory defines the $\xi_o$ as the average distance between the two electrons making up a Cooper pair determined by the uncertainty principle \cite{2016mangin}. It reads
\begin{align}\label{eq:BCS_xi}
    \xi_o = \frac{\hbar v_F}{\pi \Delta(0)}
\end{align}
where $\Delta(0) = 1.764k_BT_c$ is the zero temperature BCS gap and $v_F$ the Fermi velocity. Mayadas et al.\cite{1972mayadas} arrive at $v_F = \SI{0.62E8}{\centi\m\per\second}$ in their derivation of eq.\ref{eq:mfp-Mayadas}. Using this value we obtain $\xi_o$ between $\SI{93}{\nano\m}$ for the ex-situ film and $\SI{96}{\nano\m}$ for the in-situ - C film. 

Combing eq.\ref{eq:Bc2} and \ref{eq:xi_GL_BCS} produces
\begin{align}\label{eq:TcBc-dirty}
    B_{c2}(T) &= \frac{\phi_o}{2\pi}\frac{1}{0.7225 \ell \xi_o}\frac{T_c - T}{T_c}.
\end{align}
which expresses the upper critical field to be a linear function of temperature near $T_c$. 

Our data in fig.\ref{fig:TcBc} is indeed linear which allows us to extract the $\ell$ from eq.\ref{eq:TcBc-dirty} with $\xi_o$ given by eq.\ref{eq:BCS_xi}. The resulting mean free paths are listed in the last column of table \ref{table:comparison}. The values are close to the results from eq.\ref{eq:mfp-Mayadas} listed in table \ref{table:comparison} and follow the same trend. 

\subsection{Discussion}
Despite the observed structural changes, the critical temperatures of our Nb films vary only by a few $\SI{100}{\milli\kelvin}$. The $T_c$ of the epitaxial clean limit film from Dobrovolskiy et al. listed for comparison in table \ref{table:SputterParameters} falls within the range of our results.

The resistivities of the in-situ samples increase going down the table correlating with the deposition voltage from table \ref{table:SputterParameters}. The upper critical field increases significantly going down the table for all samples. Higher resistivities indicate structural degradation which in turn presents more pinning sites for vortices increasing the critical field \cite{2002welp,2005welp}. A significant structural change in AFM and XRD data has only been observed between the ex-situ and in-situ samples but not between in-situ samples. Thus the electrical measurements appear to be more sensitive to structural changes than the AFM and XRD analysis. 

Although the ex-situ material is smoother and has a preferred crystallite orientation perpendicular to the surface it does not have the lowest resistivity, but a lower critical field and higher critical temperatures. The superconducting properties, AFM and XRD data all indicate a more homogeneous film which should result in a lower resistivity as exemplified by the reference film by Dobrovolskiy et al. listed in table \ref{table:comparison}. An explanation could be that the ex-situ film is thinner than we expect. 

The resistivity of the clean limit reference film from Dobrovolskiy et al. is two orders of magnitude smaller despite the film being half as thick. The additional effect of surface scattering to the resistivities of thin films \cite{1938fuchs,1970mayadas} does not appear to be significant down to the $\SI{52}{\nano\m}$ thickness of the reference film. The large resitivities in our dirty limit films therefore do originate in the structural differences. 

The mean free path listed in table \ref{table:comparison} compared to the extracted coherence lengths confirms that our films are in the dirty limit. The critical field of the clean limit reference film in table \ref{table:comparison} is $\SI{0.7}{\tesla}$. Reducing the power and hence the voltage for our in-situ depositions brings us closer to the clean limit. However, at the given pressure distance of $\SI{1}{\micro\bar\m}$, we are limited to the results from in-situ - A as the lowest power that results in a stable plasma. 

\section{Nb-GaAs interface}
\begin{figure}[h]
    \centering
    \includegraphics[width = \linewidth]{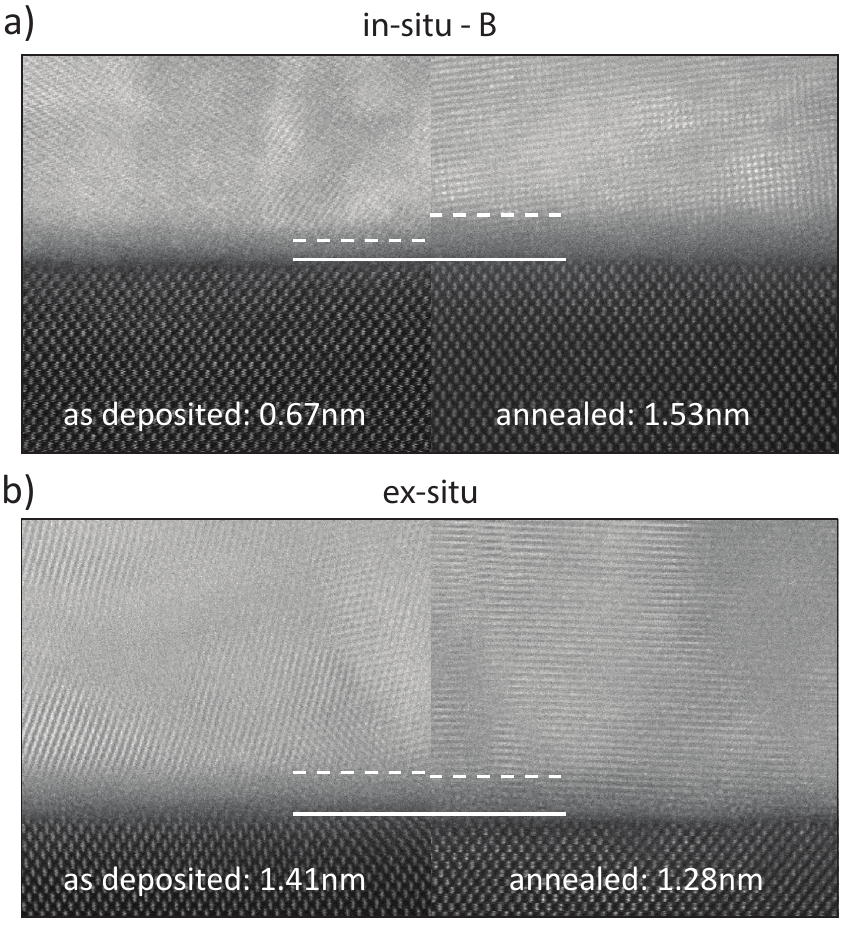}
    \caption{ADF STEM of the annealed and as deposited in-situ a) and ex-situ b) Nb-GaAs interface. See supplementary for complete images and discussion of how the interface widths were determined \cite{2023todt}.}
    \label{fig:Interlayer}
\end{figure}

ADF STEM images presented in fig.\ref{fig:Interlayer} were taken of the Nb-GaAs interface before and after annealing at $\SI{380}{\degreeCelsius}$ for $\SI{40}{\second}$ which is the annealing recipe for ohmic AuGeNi contacts to n$^{++}$ GaAs \cite{2005baca}. An amorphous interlayer can be seen for both the in-situ - B and ex-situ samples. The thicknesses of the amorphous interlayer indicated in the images were determined by comparing the normalised brightness, see supplementary for details \cite{2023todt}. The ex-situ amorphous interface, initially being thicker than in the in-situ one, is unaffected by the heat, while the in-situ material shows an increased thickness after tempering.    

The amorphous interface thickness does not increase when annealed for the ex-situ sample while for the in-situ it does by $\SI{1}{\nano\m}$. The Nb-GaAs interface is expected to be sharp up to $\SI{600}{\degreeCelsius}$ for ex-situ deposited films \cite{1988ding}. However, the quality of the TEM images presented by Ding et al. \cite{1988ding} appear to be the limiting factor in the comparison and the interlayer cannot be resolved. Nb deposition on clean GaAs surfaces via magnetron sputtering has been reported \cite{1996kutchinsky,2001giazotto} but is lacking structural investigation of the interface. 

The best reference for in-situ Nb are evaporated films on InAs nanowires, which similarly show an amorphous interlayer \cite{2017gusken,2021perla,2020carrad}. The origin of the interlayer is attributed to the formation of a Nb$_x$As$_y$ compound, which would explain our findings.

\section{Conclusion}
We have investigated both the structural and superconducting properties of Nb films in-situ deposited on GaAs in a newly designed magnetron sputtering chamber connected to an UHV MBE cluster. An exceptionally high purity of the material was achieved by designing a UHV compatible gas supply system, as proven by the residual gas analysis of the chamber and elemental analysis of the resulting Nb films.

The structural analysis via AFM of the Nb films revealed a marked difference in surface roughness between our in-situ samples and a reference ex-situ sample, deposited in a commercial system. Varying the deposition power for the in-situ samples had little effect on the surface. XRD measurements further support that difference, with the ex-situ sample having only the (110) crystallite orientation with respect to the substrate surface. The in-situ samples all showed (110), (200) and (310) oriented crystallites, despite significant differences in deposition voltages. This difference in crystallinity is attributed to the nucleation and growth of individual crystallites in the polycrystalline film. 

Measuring the superconducting resistive transition showed that the films are in the dirty limit with excellent $T_c$ and $B_{c2}(0)$ values. The $T_c$ values did not vary significantly despite at steady increase of $B_{c2}(0)$ with deposition voltage associated with structural degradation. The in-situ method produces pure films underpinned by RBS and PIXE measurements.  

The $B_{c2}(0)$ values reflected subtle structural differences between the in-situ films. With increasing deposition voltage the critical field increased, indicating that an increased power does have an effect, although we did not to resolve a trend in film structure by AFM or XRD.

All the investigated Nb - GaAs interfaces exhibited an amorphous interface layer. Tempering the samples to $\SI{380}{\degreeCelsius}$ widens the amorphous layer for the in-situ - B but not the ex-situ sample. It is unclear if the formation of the amorphous alloy between the materials is purely chemically driven or if it is related to the sputtering process. Interestingly, its presence in the in-situ deposited samples shows that it is not related to the formation of the native oxide on the semiconductor. Similar phenomena were reported in literature, and the findings indicate that it is related to formation of Nb$_x$As$_y$ alloy \cite{2017gusken,2021perla,2020carrad}.

The presented results validate in-situ magnetron sputtering as a new path to combine Nb and GaAs. Work is ongoing to employ a wider range of III-V semiconductors and superconductors to built a wider material platform for SSH. 

\section{Acknowledgement} 
The authors would like to thank Walter Bachmann, Andreas Stuker and by extension the entire workshop of the physics department of the ETH. Without their technical expertise, and more importantly patience, this system would have never been realised. Furthermore, we acknowledge Luca Alt, who built in large parts the setup used to measure the critical field and temperature and Werner Dietsche for an open ear and fruitful discussions. The project was financially supported by the Swiss National Science Foundation (SNSF). We thank the IBM Quantum Academic Network for financial support.\\
The authors gratefully acknowledge ScopeM for their support and assistance in this work, as well as the support of the clean room operations team of the Binning and Rohrer Nanotechnology Center (BRNC) and the support from Marilyne Sousa (IBM).

\bibliographystyle{apsrev}
\bibliography{references}

\pagebreak
\end{document}


\title{Supplementary: Development of Nb-GaAs based superconductor semiconductor hybrid platform by combining in-situ dc magnetron sputtering and molecular beam epitaxy}

\author{Clemens Todt}
\affiliation{Solid State Physics Laboratory, ETH Z\"urich, CH-8093 Z\"urich, Switzerland}
\author{Sjoerd Telkamp}
\affiliation{Solid State Physics Laboratory, ETH Z\"urich, CH-8093 Z\"urich, Switzerland}
\author{Filip Krizek}
\affiliation{Solid State Physics Laboratory, ETH Z\"urich, CH-8093 Z\"urich, Switzerland}
\affiliation{IBM Research Europe - Zurich, 8803 Rüschlikon, Switzerland}
\affiliation{Institute of Physics, Czech Academy of Sciences, 162 00 Prague, Czech Republic}
\author{Christian Reichl}
\affiliation{Solid State Physics Laboratory, ETH Z\"urich, CH-8093 Z\"urich, Switzerland}
\author{Mihai Gabureac}
\affiliation{Solid State Physics Laboratory, ETH Z\"urich, CH-8093 Z\"urich, Switzerland}
\author{R\"udiger Schott}
\affiliation{Solid State Physics Laboratory, ETH Z\"urich, CH-8093 Z\"urich, Switzerland}
\author{Erik Cheah}
\affiliation{Solid State Physics Laboratory, ETH Z\"urich, CH-8093 Z\"urich, Switzerland}
\author{Peng Zeng}
\affiliation{ETH Zürich, The Scientific Center for Optical and Electron Microscopy (ScopeM), CH 8093 Zürich, Switzerland.}
\author{Thomas Weber}
\affiliation{X-ray Platform, Department of Materials, ETH Zürich, Vladimir-Prelog-Weg 5–10, 8093 Zürich, Switzerland.}
\author{Arnold M\"uller}
\affiliation{Laboratory of Ion Beam Physics, ETH Zurich, Schafmattstrasse 20, CH-8093 Zurich, Switzerland.}
\author{Christof Vockenhuber}
\affiliation{Laboratory of Ion Beam Physics, ETH Zurich, Schafmattstrasse 20, CH-8093 Zurich, Switzerland.}
\author{Mohsen Bahrami Panah}
\affiliation{Solid State Physics Laboratory, ETH Z\"urich, CH-8093 Z\"urich, Switzerland}
\author{Werner Wegscheider}
\affiliation{Solid State Physics Laboratory, ETH Z\"urich, CH-8093 Z\"urich, Switzerland}

\date{\today}
\maketitle


\section{vacuum system and gas supply}
\begin{figure}[h]
    \centering
    \includegraphics{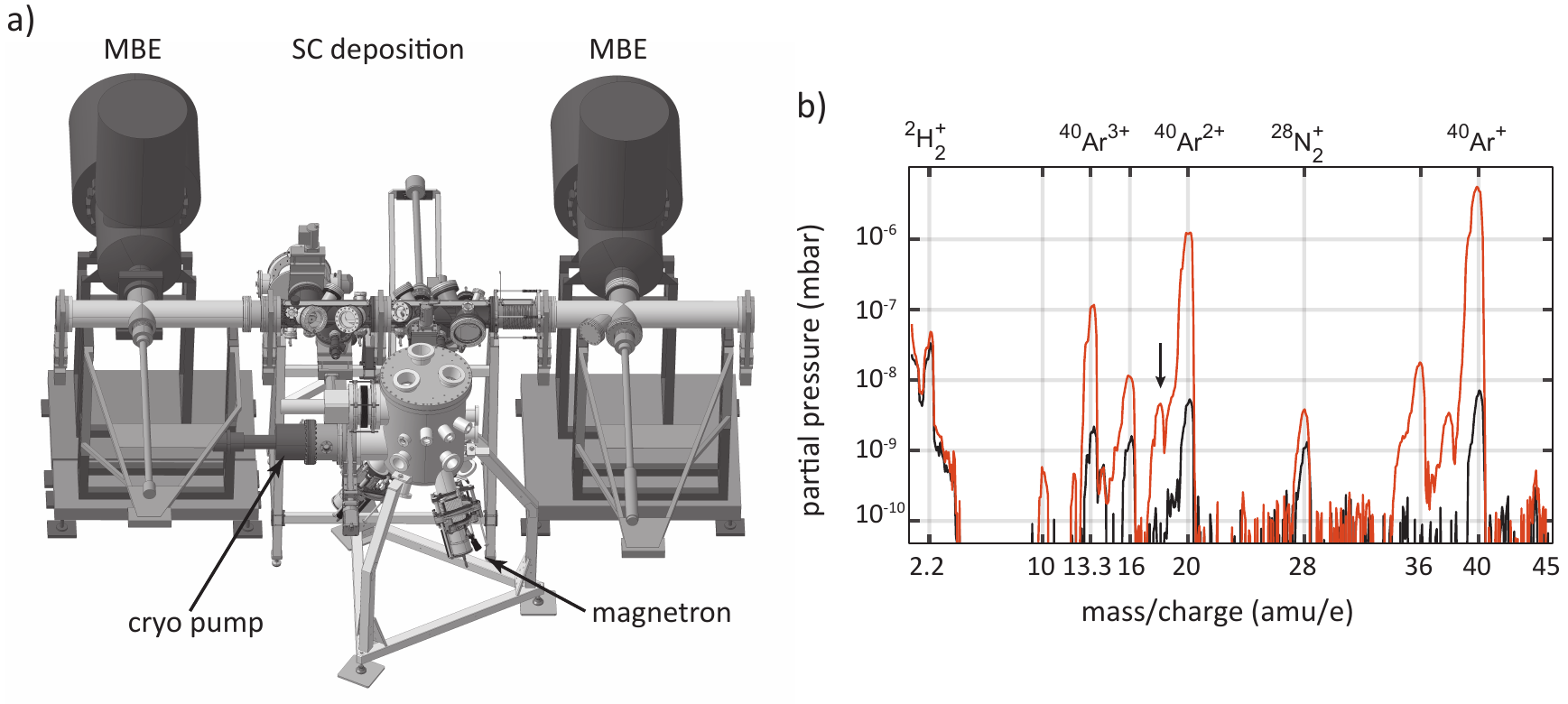}
    \caption{a) layout of the MBE and metal deposition system b) mass spectrum of the chamber. The red line indicates a measurement 18hrs after a Nb deposition while the black lines is after pumping the system for a long time.}
    \label{fig:chamber}
    \vspace{0.3cm}
\end{figure}

The magnetron sputtering system and the vacuum tunnel connecting it to the MBEs are shown in fig.\ref{fig:chamber}. Impurities known to be detrimental to the superconducting properties of elemental, A15 and B1 phase superconductors are oxygen \cite{1974koch, 2014davidhenry} and magnetic impurities such as Fe, Cr and Ni \cite{1966fulde,1978lemberger}. Attention must therefore be paid to remove adsorbed H$_2$O and CO$_2$ from the chamber and gas supply. Magnetic impurities occur as solids and may originate either in the source material or from resputtering of steel parts. Hydrocarbons are a third contaminant whose influence on the superconducting film is unknown and originate from rubber seals, lubricants and residual solvents from degreasing. 

The entire system is built to be ultra high vacuum capable meaning that all seals are made of metal and the magnets in the magnetrons are removable without breaking vacuum. This allowed us to bake the system past the Curie temperature of the magnets to $> \SI{150}{\degreeCelsius}$ removing H$_2$O and CO$_2$ adsorbed on the chamber surfaces.

Moving parts in vacuum are designed to work without a lubricant and mechanical feedthroughs are magnetically coupled. The valve seals are made from the fluorocarbon elastomer FKM with an excellent high temperature and low pressure stability.

Magnetron sputtering as opposed to evaporation requires gas flow into the chamber which potentially carries contaminants. Commercially available Argon comes in 6N ($\SI{99.9999}{\percent}$) purity and will carry with it contamination from the gas lines. Therefore gas purifiers were place close to the system. The Argon gas is washed using the SAES GC-50 getter furnace which removes H$_2$O, O$_2$, CO, CO$_2$ and H$_2$ to $<\SI{2e-7}{\atpercent}$. The getter also removes N$_2$ unwanted for non reactive sputtering. The gaslines themselves are all metal constructions with swagelok assembly by torque fittings connecting the bottle to the gas washer. The metal gaslines are made from swagelok VCR metal gasket face seal fittings from the getter furnace to the mass flow controller and ultimately the chamber, providing the highest leak integrity. After assembly and flushing the gaslines were baked prior to conditioning the main chamber thus removing specifically water from the system. The entire gas system up to the mass flow controllers is kept at a slight over pressure with respect to air. The mass flow controllers are metal sealed GM50A from MKS with a external leak integrity of $<\num{1e-9}\nicefrac{scc}{sec}$ of He.

The system is pumped by a Leybold CoolVac 1500 customized to feature only UHV compatible materials on the vacuum side. A cryo pump is particularly efficient in removing contaminate pumping H$_2$O $\SI{4600}{\liter\per\second}$. However the Ar pumping speed through the DN200CF flange is far too large at $\SI{1200}{\liter\per\second}$ to be useful in magnetron sputtering. Hence a bypass with a custom UHV control valve allows reasonable pressure regulation in down stream control setup. In the dormant state the system is pumped through the DN200 flange resulting in an excellent vacuum quality. Additionally the vacuum quality in the chamber naturally improves with usage, since Nb deposited on the walls is an excellent getter \cite{2010mattox,1992henriot}. 

In fig.\ref{fig:chamber} from the main text is reproduced where a) shows a schematic of the system and b) are typical mass spectra recorded before and after magnetron sputtering of films. The main gas present is naturally argon with the main peak at 40 and the higher ionized states at 20 and 13.3. A small peak associated with nitrogen appears at 28 which could be the remnant impurity not removed by the getter furnace. The peak at 16 is unidentified and the shoulder at 19 is associated with F$^+$ the origin of which is equally unknown. We can exclude CH$_4^+$ since CH$_3^+$ (15), CH$_2^+$ (14) and CH$^+$ (13) are missing, O$^+$ since O$_2^+$ (32) and OH$^+$ (17) are missing as well as CO$^+$ since CO$_2^+$ (44) is not present. The most concerning although be it tiny signal is the water peak at 18 after sputtering. The gas system is optimized to remove water and keep it out. With time and usage the water signal disappeared entirely. 
\clearpage
\section{elemental analysis}
The elemental composition of the ex-situ and in-situ samples was investigated with Rutherford Back Scattering (RBS) using $\SI{2}{MeV}$ He$^+$. The particle induced x-ray emission (PIXE) was measured in tandem. The aim was to determine any significant impurities and information on the width and composition of the Nb - GaAs interface. The 4N4 Nb targets for both ex-situ and in-situ samples were supplied by AJA Int. with the main residual elements being Ta(150ppm), C(79ppm), O(71ppm), N(40ppm) and Mo(24ppm). 

\begin{figure}[h]
    \centering
    \includegraphics{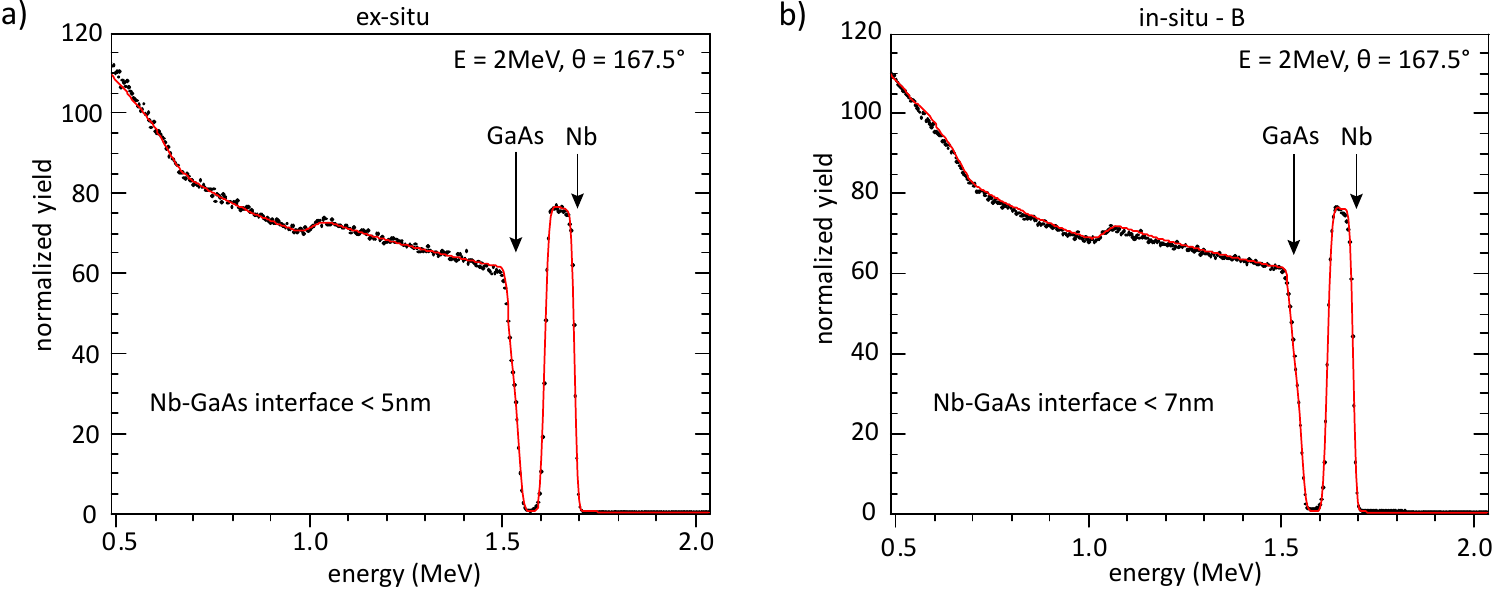}
    \caption{Rutherford Back Scattering data using $\SI{2}{MeV}$ He$^+$ for a) ex-situ and b) in-situ - B samples }
    \label{fig:RBS}
    \vspace{0.5cm}
\end{figure}

The RBS data in fig.\ref{fig:RBS} shows the expected signal of Nb and the GaAs substrate. The information of the interface width is limited to the depth resolution of the RBS measurement and can only give us the upper bound of $\SI{5}{\nano\m}$ and $\SI{7}{\nano\m}$ for the ex-situ and in-situ - B samples. The RBS signal increases as $\nicefrac{1}{E^2}$ \cite{2015nastasi} towards lower energies obscuring the expected signal form the surface oxide \cite{1991chevarier}. 

\begin{figure}[h]
    \centering
    \includegraphics{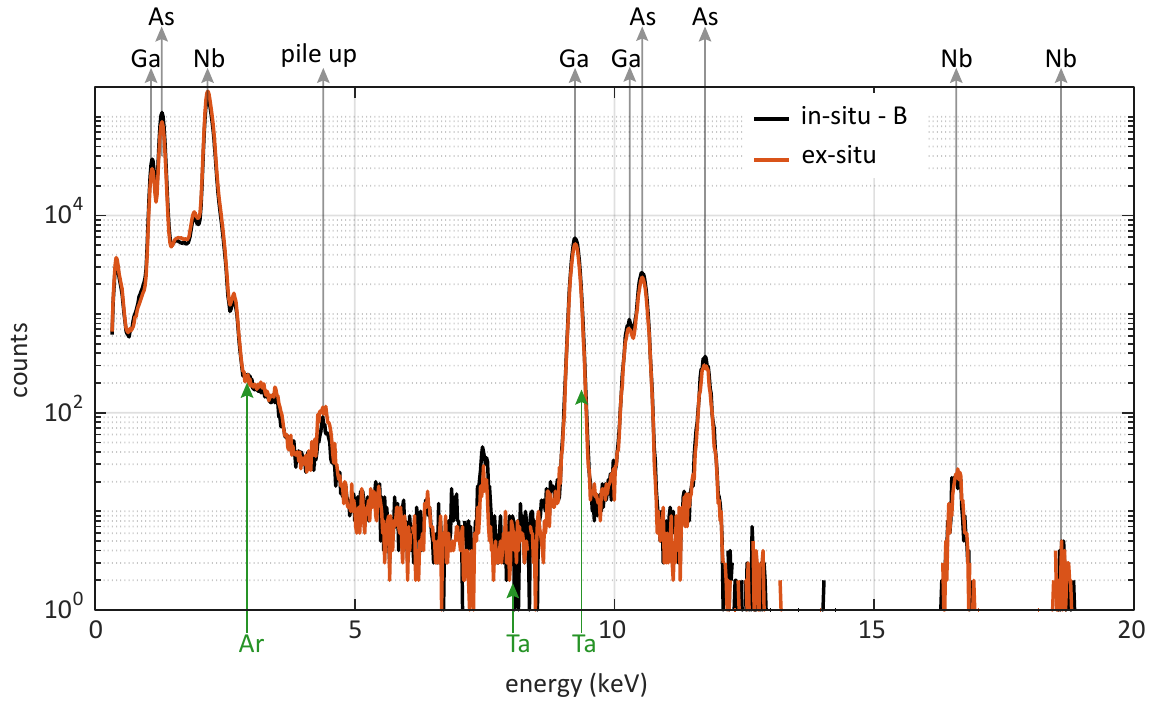}
    \caption{particle induced X-ray emission data using $\SI{2}{MeV}$ He$^+$ obtained for the ex-situ and in-situ - B samples.}
    \label{fig:PIXE}
\end{figure}

The PIXE data given in fig.\ref{fig:PIXE} complements the RBS findings. The expected peaks for As, Ga and Nb are indicated. 

Argon incorporation into the film has been identified as causing compressive strain \cite{1993window,1999iosad}. The expected principal Ar peak as indicated in fig.\ref{fig:PIXE} does not appear, which allows us to estimate the Ar incorporation for both films to be $<\num{0.001}$ with respect to the Nb signal. 

According to the supplier Ta is the main target contamination at (150ppm). A presence of Ta in the PIXE spectrum would be signified by peaks at $\SI{9.3}{\kilo\electronvolt}$ and $\SI{8.1}{\kilo\electronvolt}$, which have been indicated in fig.\ref{fig:PIXE}. While the higher energy line is likely obscured by the Ga signal the lower energy signal should appear, if a large enough amount of Ta is present. The relative intensities between the higher and lower X-ray energies are $\num{0.8094}$ and $\num{0.7185}$ \cite{2015nastasi}, respectively.

The signal at $\SI{7.48}{\kilo\electronvolt}$ is likely an artefact and not due to Ni. The K-$\beta$ at $\SI{8.26}{\kilo\electronvolt}$ is missing and if Ni was an impurity in the Nb film it would indicate a contamination of $\num{2.(5)}\si{\atpercent}$. Such a large signal would not be explainable by the quoted Ni contamination of 0.05ppm in the target, nor can we identify another source of Ni during the deposition or handling of the samples. If a significant contamination of a magnetic impurity of Ni was present in the Nb, a reduction of critical temperature, and a broadening of the resistive transition would be expected\cite{1966maki,1978lemberger}. However our measurements in fig.\ref{fig:TcSweeps} and fig.\ref{fig:BcSweeps} do not show such an effect. Therefore the Ni signal is regraded as measurement artefact. 

\clearpage
\section{electrical measurement}
The first set of van der Pauw samples (ex-situ and in-situ - B) were lithography defined dots $\SI{200}{\micro\m}$ in diameter. The contacts were $\SI{10}{\micro\m}$ wide Nb leads with evaporated Au contacts. The second generation of samples (in-situ - A and in-situ - C) were $\SI{5}{\milli\m}$ by $\SI{5}{\milli\m}$ cleaved pieces with In contacts on the corners which are faster to produce and resulted in more stable measurements. 

A lock-in amplifier (Zurich Instruments MFLI) provided the oscillator voltage $V_{osc}$ that was applied to a bias resistor of $\SI{1}{\kilo\ohm}$ and the resulting current was fed to the sample in a liquid Helium fridge. The current was sunk and measured at a second lock-in, while the first lock-in picks up the differential voltage, dropped in parallel to the current flow. To obtain a reliably measurable voltage signal, a root mean square current of $\SI{108}{\micro\ampere}$ was necessary at a measurement frequency was $\SI{1.711}{\kilo\hertz}$. This frequency gave a minimum in noise and did not yet shown a change amplitude or phase down to low frequencies. 

The temperature range of the fridge is $\SI{4.2}{\kelvin}$ to $\SI{30}{\kelvin}$ with a maximum inaccuracy in the measured temperature of $\SI{50}{\milli\kelvin}$. The magnet can provide magnetic field up to $\SI{3}{\tesla}$ perpendicular to the sample surface.

\begin{figure}[h]
    \centering
    \includegraphics{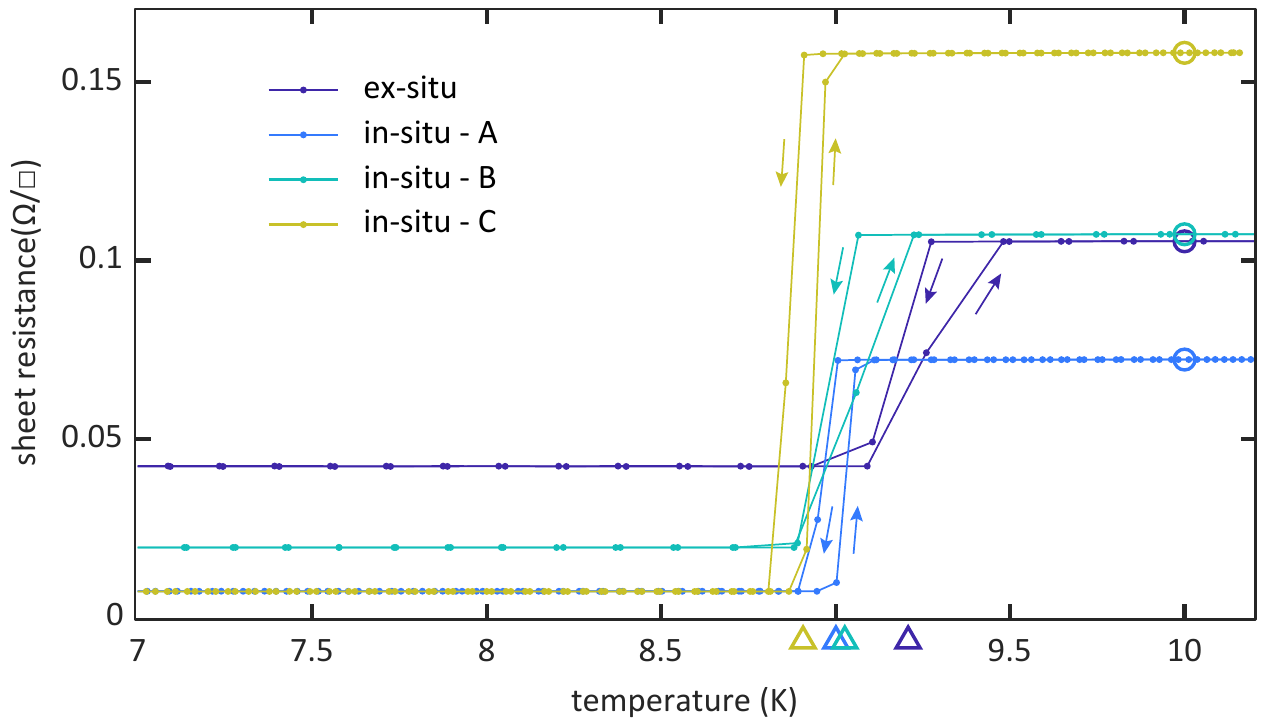}
    \caption{measured temperature dependence of the sheet resistance. The extracted normal state sheet resistance at $\SI{10}{\kelvin}$ is indicated by the circles. The critical temperatures are indicated on the temperature axis. The arrows indicate sweep direction.}
    \label{fig:TcSweeps}
\end{figure}

The measured resistance at zero field versus temperature is given in fig.\ref{fig:TcSweeps}. The temperature was raised from $\SI{4.2}{\kelvin}$ to above $\SI{10}{\kelvin}$ and then lowered again to $\SI{4.2}{\kelvin}$ to reveal a possible hysteresis. The difference between normal state and superconducting is quite small due to the low sample resistance of $<\SI{1}{\ohm}$. In the superconducting state the lock-in can no longer demodulate a signal and the phase of the measured quantities becomes random. Therefore the remnant signal below $T_c$ is due to noise.  

\begin{figure}[h]
    \centering
    \includegraphics{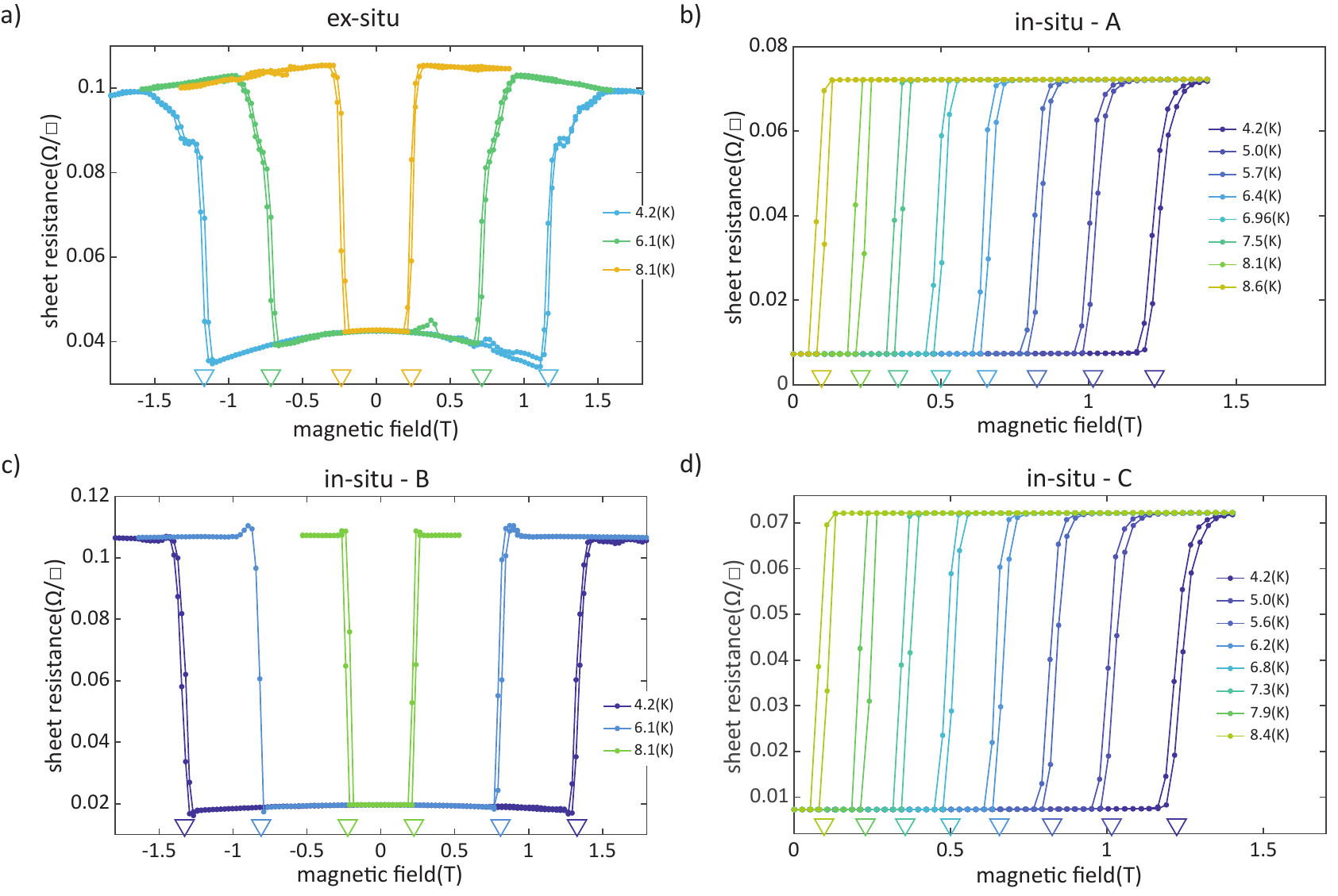}
    \caption{measured perpendicular magnetic field dependence of the sheet resistance at varying temperatures. The extracted critical temperatures are indicated on the magnetic field axis. The samples are a) ex-situ b) in-situ - A c) in-situ - B d) in-situ - C }
    \label{fig:BcSweeps}
\end{figure}

The critical field measured at varying temperatures is shown in fig.\ref{fig:BcSweeps} for all four samples. The extracted critical magnetic fields are indicated on the magnetic field axis. The ex-situ and in-situ samples were measured for negative and positive values. It was established that the critical field value does not depend on the polarity of the applied magnetic field. All measurements were done sweeping the field up and back down to exclude hysteresis effects. 

The lithography defined samples (ex-situ) and (in-situ - B) display several odd features. The ex-situ sample has an overall bow to it and displays spurious signal increase in the superconducting state. The in-situ - C sample has a small peak just before the transition which is most pronounced in an intermediate temperature of $\SI{6.1}{\kelvin}$. The resistive temperature superconducting transitions were nonetheless clearly visible. The cleaved square van der Pauw pieces of samples in-situ - A and in-situ - C did not show any additional features and had the lowest noise floor. 

\clearpage
\section{TEM extraction}
\begin{figure}[!h]
    \centering
    \includegraphics{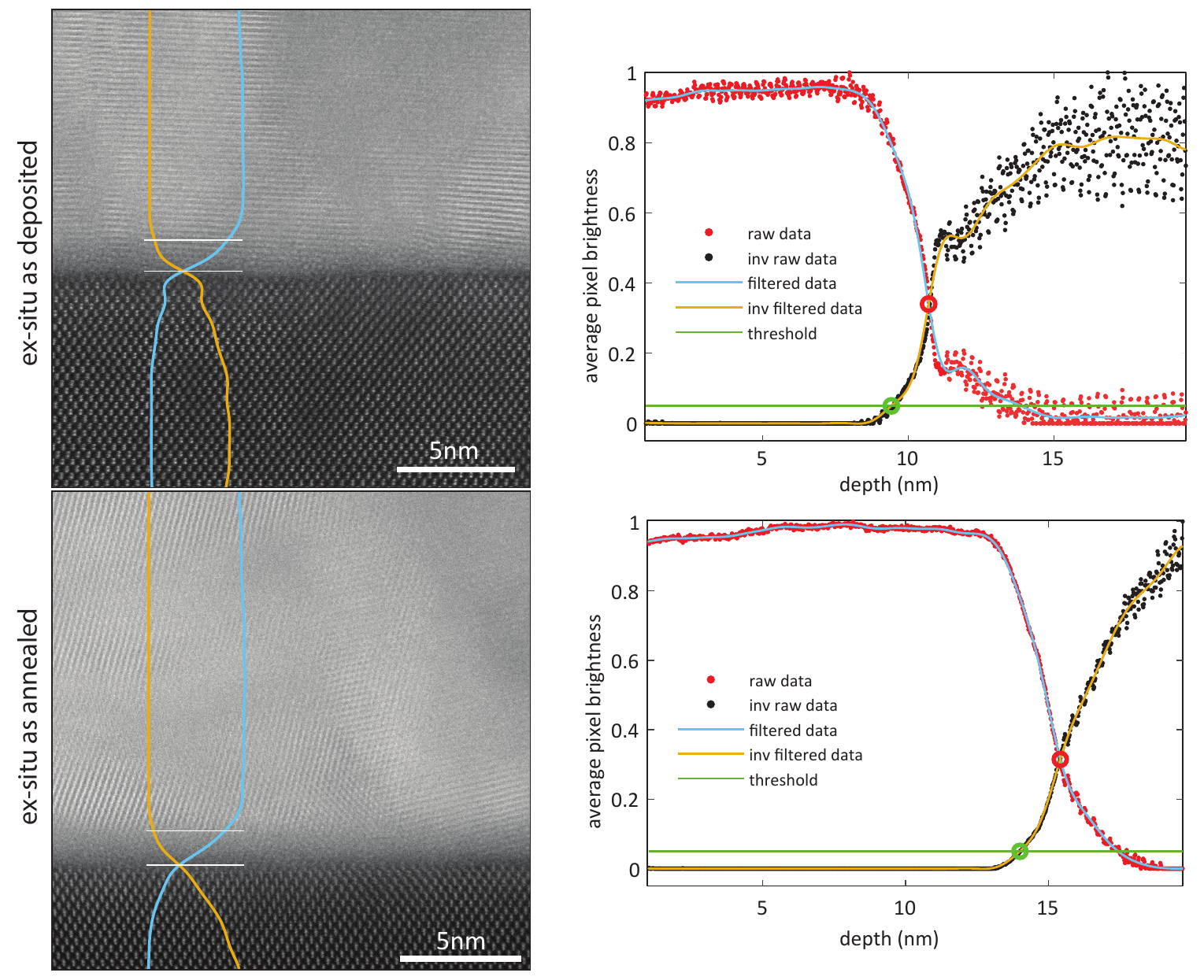}
    \caption{complete annular dark field TEM images of the ex-situ Nb to GaAs interface as presented in the main text. The analysis data for the extraction of the amorphous interlayer width it show next to it.  }
    \label{fig:TEMInterlayers ex-situ}
\end{figure}

\begin{figure}[!h]
    \centering
    \includegraphics{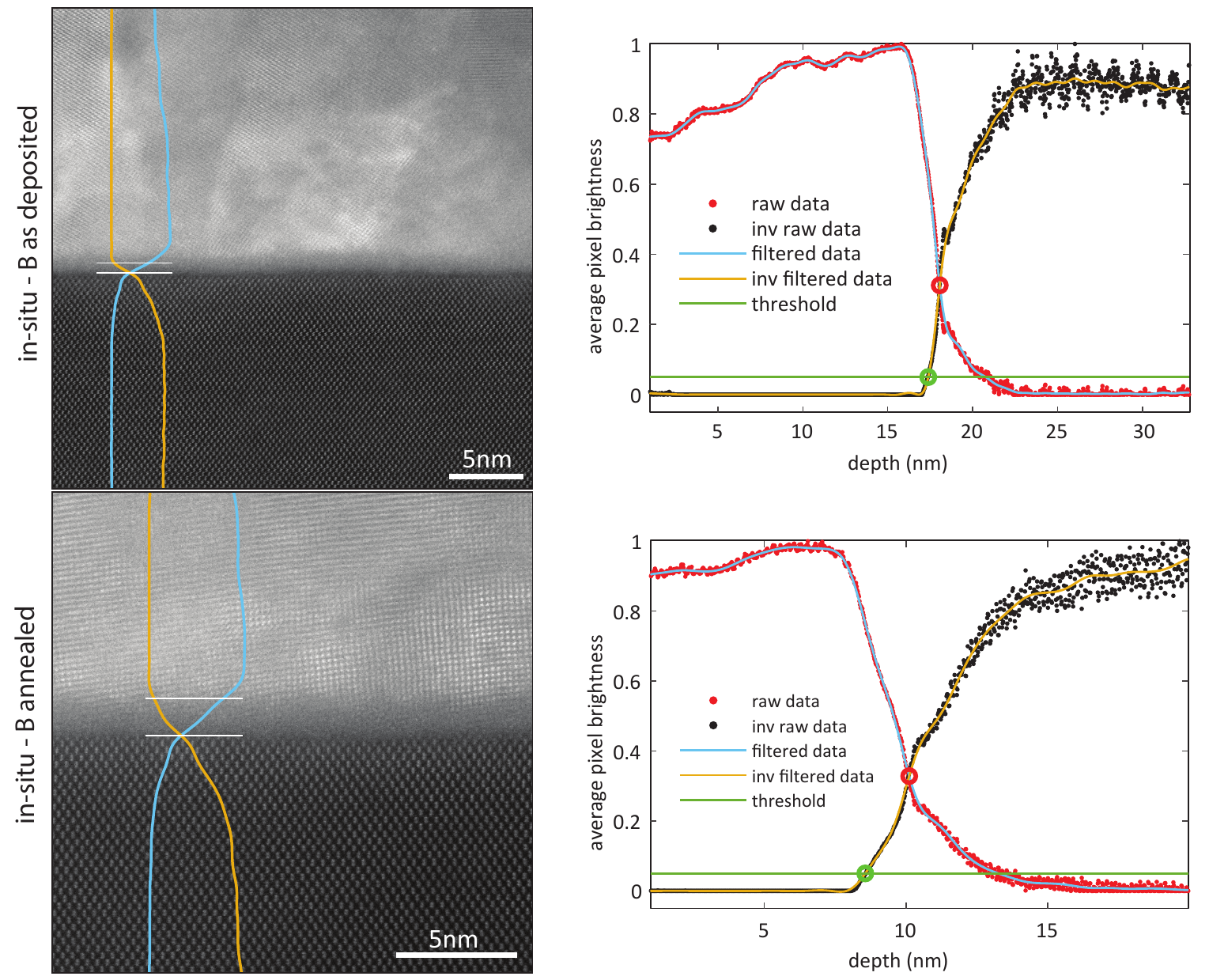}
    \caption{complete annular dark field TEM images of the in-situ - B Nb to GaAs interface as presented in the main text. The analysis data for the extraction of the amorphous interlayer width it show next to it.  }
    \label{fig:TEMInterlayers in-situ}
\end{figure}

The width of the amorphous interface was numerically determined from annular dark field transmission electron microscope image presented in fig.\ref{fig:TEMInterlayers in-situ} and fig.\ref{fig:TEMInterlayers ex-situ}. The crystalline GaAs substrate appears darker with a perfect crystallographic structure while the Nb gives a bright signal with randomly oriented grains. The amorphous interlayer appears less bright than the Nb itself.

All images were rotated such that the GaAs substrate aligned with the bottom of the image. The pixel brightness of each row was averaged and normalised to one, resulting in the data given next to each ADF TEM image in fig.\ref{fig:TEMInterlayers in-situ} and fig.\ref{fig:TEMInterlayers ex-situ}. The data was low pass filtered to reduce noise.  The averaged pixel brightness for each row is shown by red dots with the filtered data as bright blue lines. The inverse average pixel brightness is given by the black dots with the filtered data shown in orange.  

The GaAs surface was detected by finding the intersection between the average pixel brightness and its inverse indicated by a red circle. Determining the end of the amorphous interface towards the Nb requires an arbitrary threshold we chose to be 0.05. In order to achieve comparability the threshold was kept constant between images. The amorphous interlayer widths we thus determined are indicated by a green circle. 

\clearpage
\bibliographystyle{apsrev}
\bibliography{references}

\pagebreak